# Accelerating inverse crystal structure prediction by machine learning: a case study of carbon allotropes

Wen Tong[a], Qun Wei[a,*], Haiyan Yan[b], Meiguang Zhang[c,**] Xuanmin Zhu[d]

[a] *School of Physics and Optoelectronic Engineering, Xidian University, Xi'an 710071, China*
[b] *College of Chemistry and Chemical Engineering, Baoji University of Arts and Sciences, Baoji 721013, China*
[c] *College of Physics and Optoelectronic Technology, Baoji University of Arts and Sciences, 721016 Baoji, China*
[d] *School of Information, Guizhou University of Finance and Economics, Guiyang, 550025, China*

**Abstract**

Based on structure prediction method, the machine learning method is used instead of the density function theory (DFT) method to predict the material properties, thereby accelerating the material search process. In this paper, we established a data set of carbon materials by high-throughput calculation with available carbon structures obtained from the Samara Carbon Allotrope Database. We then trained an ML model that specifically predicts the elastic modulus (bulk modulus, shear modulus, and the Young's modulus) and confirmed that the accuracy is better than that of AFLOW-ML in predicting the elastic modulus of a carbon allotrope. We further combined our ML model with the CALYPSO code to search for new carbon structures with a high Young's modulus. A new carbon allotrope not included in the Samara Carbon Allotrope Database, named *Cmcm*-C24, which exhibits a hardness greater than 80 GPa, was firstly revealed. The *Cmcm*-C24 phase was identified as a semiconductor with a direct bandgap. The structural stability, elastic modulus, and electronic properties of the new carbon allotrope were systematically studied, and the obtained results demonstrate the feasibility of ML methods accelerating the material search process.

[*] Corresponding author,  [**] Corresponding author.
*E-mail addresses:* weiaqun@163.com (Q. Wei). zhmgbj@126.com (M. Zhang).

# 1. Introduction

Machine learning (ML), as a data-driven approach for forecasting, decision making or classification, is playing an increasingly important role in materials science. For example, Deringer et al. [1] confirmed that a novel class of ML-based interatomic potentials can be used for random structure searching and to readily predict several hitherto unknown carbon allotropes. Zhuo et al. [2] developed an ML model that can accurately predict the bandgaps of inorganic solids. For the binary semiconductors [3] of groups III–V, II–VI and the ternary semiconductors [4] of groups I–III–VI2 and II–IV–V2, the accuracy of bandgap predictions using a neural network model was similar to that of predictions using density functional theory (DFT). In addition, support vector machine (SVM) models have been used to classify the crystal structures of transition metal phosphides and to predict equiatomic binary compounds leading to the discovery of several novel phases [5, 6]. Using the Inorganic Crystal Structure Database, Legrain et al. [7] trained an ML model to predict the vibration free energy and entropy of the compound crystals based on the material descriptors and found that the prediction accuracy was similar to that of DFT calculations. In 2016, Pilania et al. [8] established an ML model for the perovskite structure and successfully predicted a new perovskite structure. In 2017, Isayev et al. [9] used a combination of first-principles data and material structure relationship models in a materials database to establish an ML model that could quickly predict the properties of inorganic materials. To solve the problem of an insufficient material database of the target type, Zhang et al. [10] proposed a strategy of training a generalized ML model on a small data set, thereby expanding the application range of the ML method.

A computer-assisted inverse-design method for searching a special functional materials has been recently developed [11–13]. The main goal of this method is not only to find the global and/or local minima of the free energy surface but also to screen out structures with certain properties. The CALYPSO structure prediction method based on swarm intelligence is widely used in structural searches for new materials [14], and this method has been successfully applied to various systems [15–20]. Although this method has achieved wide success, the structures produced by CALYPSO in each generation have been optimized by first-principles calculations, thus costing valuable computing resources. In 2018, Tong et al. [21] used the method of constructing the ML potential to accelerate the prediction of crystal structures. Xia et al. [22] used a Gaussian process model to predict the energies of structures in the structure search process. After the structural optimization using ML, the calculation speed of the structural optimization part was greatly improved. However, in the process of structure inverse-design, the most time-consuming step is the property calculation of each crystal structure. Improving the efficiency of property predictions is a major challenge.

In the present work, on the basis of ML, we attempt to accelerate the calculations of elastic modulus. First, we trained three ML models for predicting the elastic modulus of known carbon materials using random forest (RF) algorithm, support vector machine regression (SVR) algorithm, and an artificial deep neural network (DNN) algorithm. After comparing the prediction accuracy of these three models, we chose the RF model as our final model. In addition, we confirmed that the accuracy of the RF model for predicting the elastic moduli of carbon materials is better than that of AFLOW-ML. For each generation of the structures generated by CALYPSO, we used ML models instead of DFT calculations to predict the elastic modulus of the structures, thereby accelerating the search process for a novel carbon structure with high elastic modulus. Ultimately, we obtained a new superhard carbon allotrope not included in the SACADA database [23]. The structural stability, elastic modulus, and electrical properties of this carbon allotrope were systematically studied.

## 2. Materials and methods

2.1 Computational method

The crystal structure prediction is based on the global minimization of energy surfaces merging ab initio total energy calculations as implemented in the CALYPSO code [14].

The variable cell structure predictions were performed at 0 GPa with 1 to 40 carbon atoms per simulation cell, and the structure searching simulation was stopped after ~5000 structures generated (the number of structures of each generation was set to 100). All calculations based on the first-principles method of density functional theory (DFT) and the projector augmented wave (PAW) [24] were performed via the Vienna ab initio simulation package (VASP) [25, 26]. The Perdew–Burke–Ernzerhof (PBE) functional of the generalized gradient approximation (GGA) [27, 28] was employed as the exchange and correlation functional. The cutoff energy of 900 eV were set for expanding the wave functions. The convergence criteria for structural optimizations were that the total energy of two adjacent steps should be less than $10^{-5}$ eV and the force per atom in the structural unit should be less than 0.05 eV/Å. The phonon spectra were calculated by using a finite displacement approach implemented in the PHONOPY code [29].

2.2 ML accelerated crystal structure searching

We proposed and implemented a method based on ML to accelerate crystal structure prediction and improve the search efficiency. The complete algorithmic process is shown in Figure 1. An ML model that predicts specific properties was used to replace the property calculations based on VASP. We applied the improved structural search scheme to carbon materials. In the structure search process, we set the structural parameter that needed to be predicted as the elastic modulus, and finally obtained a new carbon phase that has not been previously reported.

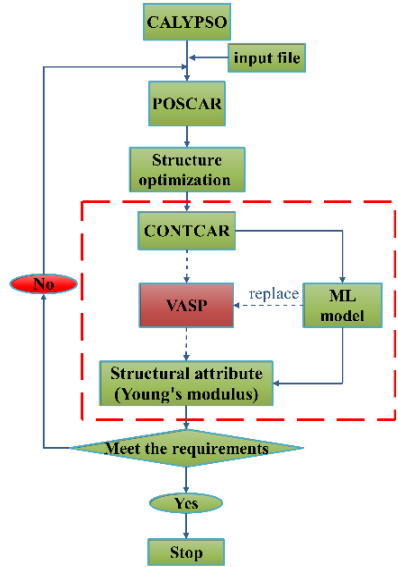

Figure 1. The flow chart of the machine-learning-accelerated crystal structure prediction method.

2.3 Data extraction and ML training model.

SACADA is a database containing many available data about three-dimensional (3D) carbon allotropes [23]. To supply the data set for ML, the crystal structures of 522 carbon materials extracted from the SACADA database were fully reoptimized on the basis of DFT using VASP within the GGA–PBE level. As a result, crystals with a high-pressure phase or unreasonable structure after optimization were removed and 490 reasonable crystal structures were retained; the elastic constants and bulk and shear moduli ($B$ and $G$) of the retained structures were calculated by DFT, and the Young's modulus ($E$) was calculated by the formula [20] $E = 9BG/(3B+G)$. In the case of a small data set of carbon materials, choosing the feature descriptor with a higher correlation with the elastic modulus can avoid underfitting and improve generalization in the ML model. The MATMINER [30], a Python-based third-party library focused on material feature mining, can mine 273 descriptors used to describe crystal structures. In addition, adding irrelevant descriptors into the feature space will increase the complexity of the ML model and lead to overfitting of the model. An efficient method for screening out irrelevant features is to calculate the Pearson product-moment correlation coefficient (PPMCC) between the feature space and the target attribute (i.e. elastic modulus). PPMCC is defined as the quotient of covariance and standard deviation between two variables, and the coefficients range from −1 to 1. After standardizing the data(where standardization refers to the data minus its average and then divided by its standard deviation to obtain data with a standard normal distribution with an average of 0 and a standard deviation of 1), 15 structure descriptors related to variables including the crystal system, space group, and the unit-cell volume, among others, were used to train and test the ML model. Among them, the PPMCC between density and elastic modulus was as high as 0.92. The extracted crystal structure descriptors constitute the feature space in the data set, and the elastic modulus was used as the target attribute in the data set.

The data set was divided into a training set and a test set. Because the data set was too small, the experiment was based on the distribution of the density descriptor values of the carbon structures in the dataset; 80% of the data sets were selected as the training set and 20% were used as the test set. The test-set and training-set partitioning is shown in Figure 2, where the carbon structures in the test and training sets are shown to follow the same distribution.

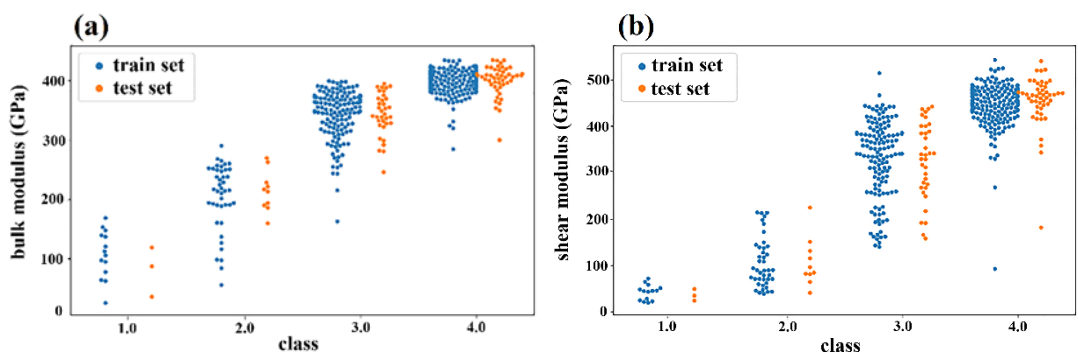

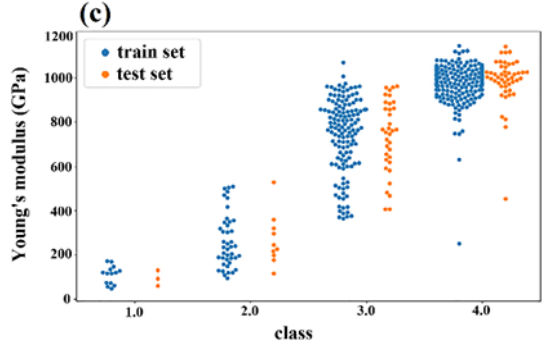

Figure 2. "Class" indicates the four data sets into which the data set were divided according to the density distribution of the crystal structure. Twenty percent of each set was set as the test set, and 80% was set as the training set; the vertical axis reflects the distribution of bulk modulus (a), shear modulus (b), and Young's modulus (c) in the training set and the test set.

Here, we selected the RF algorithm, SVR algorithm and DNN algorithm as the training algorithms for the ML model with a $k$-fold cross validation scheme, which refers to dividing the training set into $k$ sample subsets, each time using one sample subset as the validation set and the remaining $k - 1$ sample subsets as the training set, repeated $k$ times. The cross-validation method helps to avoid overfitting and artificially high statistics. Following the RF, DNN, and SVR methods, 490 samples were used to train and test models that predicted $B$, $G$, and $E$, respectively. Excellent agreement was obtained between the DFT-calculated $B$, $G$, and $E$ values and the ML-predicted values, with the cross-validated root-mean-square error (CV-RMSE) following $\frac{1}{k}\sum_k(\sqrt{\frac{1}{n}\sum_{i=1}^{n}(y_i - \hat{y}_i)^2})$, where $k$ is the number of sample subsets, $cv = 5$ in the present paper, $n$ is the size of sub-sample set, and $y_i$ and $\hat{y}_i$ are the original and predicted values, respectively. CV-RMSE is commonly used and it makes an excellent general-purpose error metric for numerical predictions. The CV-RMSE values of the models trained by different algorithms are shown in Figure 3. In the predictions of the shear modulus, bulk modulus, and the Young's modulus, the performance of the RF model was better than those of the DNN model and the SVR model. Thus, under the condition of a small data set, the integrated learning method can effectively reduce the error of the model. We finally selected the RF model as a predictive model of the elastic modulus of carbon materials.

The RF model was trained according to Scikit-Learn, an open source third-party Python ML library. In the RF model, the main optimized hyperparameters were the number of decision stumps that make up the RF (n_estimators) and the maximum number of features used by the decision tree (max_features). The n_estimator and max_features of the RF model used to predict the Young's modulus were 123 and 8, respectively. The n_estimator and max_features of the RF model used to predict the shear modulus were 41 and 10, respectively. The n_estimator and max_features of the RF model used to predict the bulk modulus were 101 and 6, respectively.

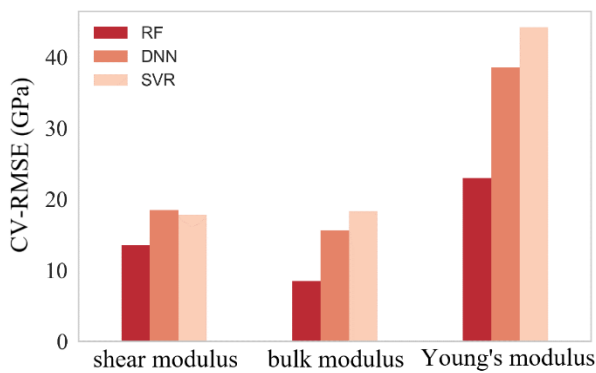

Figure 3. Cross-validated root-mean-square error (CV-RMSE) of the predictions of the shear modulus, bulk modulus, and Young's modulus using the random forest (RF), deep neural network (DNN), and support vector machine regression (SVR) models.

## 3 Results and discussion

Our trained model and AFLOW-ML [31-33] predicted 490 carbon structures, the results are shown in Figure 4. In Figure 4(a), the RF model predicts the bulk modulus with an RMSE of 10.16 GPa, which is 21.10 GPa less than the RMSE of the value predicted by AFLOW-ML. In Figure 4 (b), the RF model predicts a shear modulus with an RMSE of 15.92 GPa, which is a 39.23 GPa lower than the RMSE of the prediction of AFLOW-ML. For the predictions of the Young's modulus, because it cannot be obtained by AFLOW-ML directly, we used the formula $E = 9BG/(3B+G)$ (where $B$ and $G$ are obtained by AFLOW-ML) to represent the AFLOW-ML results. Figure 4(c) shows that the RMSE for the RF prediction of the Young's modulus is 31.11 GPa, which is 73.60 GPa lower than the RMSE of the AFLOW-ML direct prediction. These results show that the model error of predicting the Young's modulus is greatly reduced compared with the results of AFLOW-ML. In the RF model, the error only arises from the DFT calculation and the error of the model itself. In AFLOW-ML, the Young's modulus calculated using the values of the bulk modulus and shear modulus predicted by AFLOW-ML (since AFLOW-ML cannot predict the Young's modulus) leads to the error including the error of the AFLOW-ML-predicted bulk modulus and shear modulus.

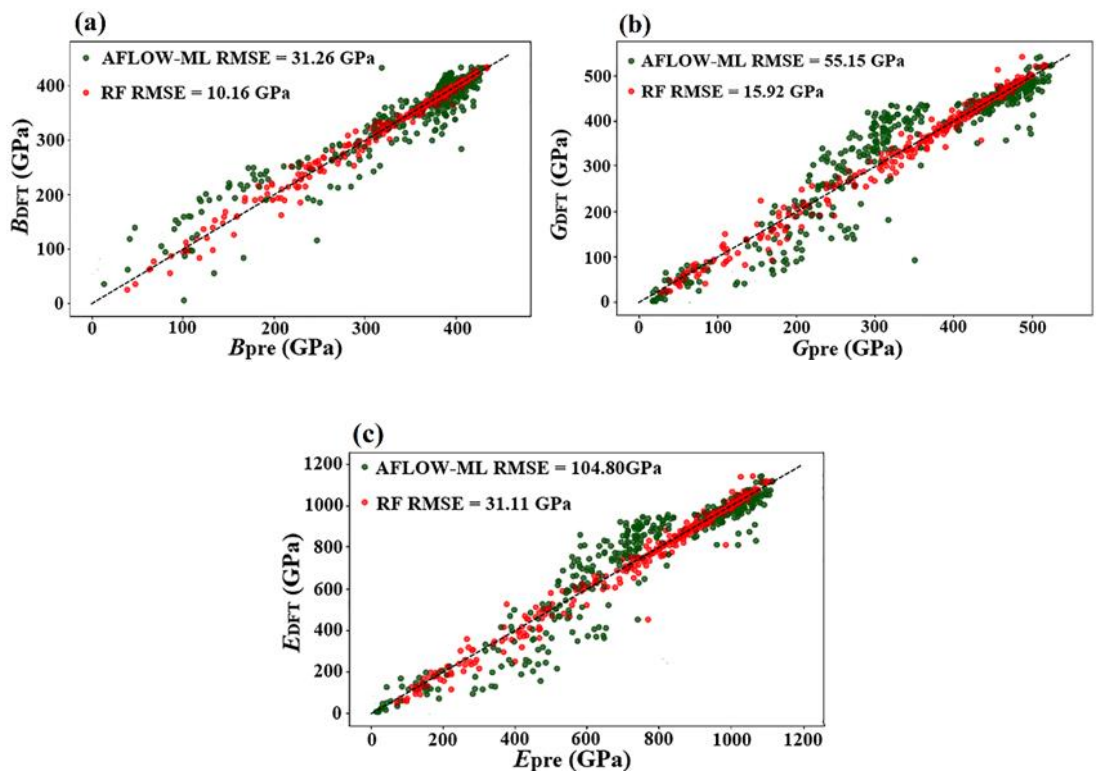

Figure 4. Cross-validated values of the bulk modulus $B$ (a), shear modulus $G$ (b), and Young's modulus $E$ (c) using the RF and AFLOW-ML models against the data set. The red dots in the figures indicate the prediction results of the RF model, and the green dots indicates the prediction results of AFLOW-ML; the dashed line indicates the reference line, and the closer a point is to the dotted line, the more accurate the prediction result.

We added the model of predicting the Young's modulus to the structure search processing and searched a large number of stable structures within a short time. Numerous well-known superhard carbon allotropes with a high Young's modulus, including lonsdaleite, *M*-carbon, *Z*-carbon, and *F*-carbon. [34-38], have also been reproduced, which demonstrates that our method is feasible. In addition, we found a stable superhard carbon phase with relative enthalpy ($\Delta H$) to diamond less than 0.40 eV/atom. The crystal structure of this new carbon phase belongs to the space group *Cmcm* (space group no. 63). A unit cell comprises 24 carbon atoms. The lattice parameters of this *Cmcm*-C24 are $a$ = 7.641 Å, $b$ = 4.361 Å, and $c$ = 4.459 Å. The *Cmcm*-C24 has 16*h*, and 8*f* Wyckoff positions: C1 16*h* (−0.67397, −0.34074, 0.56989), and C2 8*f* (0.00000, −0.34751, 1.42327). The crystal structure of *Cmcm*-C24 is shown in Figure 5(a). The enthalpy of the carbon phase relative to diamond was calculated using the formula $\Delta H = H_{\text{new phase}}/n_1 - H_{diamond}/n_2$, where $n_1$ and $n_2$ are the number of atoms per unit cell of each carbon phase. The calculated relative enthalpy is 0.32 eV/atom, as listed in Table 1. The elastic constants of *Cmcm*-C24 are $C_{11}$ = 928 GPa, $C_{22}$ = 1039 GPa, $C_{33}$ = 738 GPa, $C_{44}$ = 397 GPa, $C_{55}$ = 314 GPa, $C_{66}$ = 444 GPa, $C_{12}$ = 98 GPa, $C_{13}$ = 142 GPa, and $C_{23}$ = 125 GPa. Clearly, the elastic matrix is positive definite, which means *Cmcm*-C24 is mechanically stable. The shear modulus, Young's modulus, bulk modulus, and Poisson's ratio were obtained by using the Voigt-Reuss-Hill approximation [39] and are shown in Table 1. The 3D directional dependence of the Young's modulus of *Cmcm*-C24 is shown in Figure 5(b). As shown in Figure 5(b), the new structure exhibits substantial anisotropy, as indicated by its contour lines of the Young's modulus in the *xy*, *yz*, and *zx* planes deviating from a circle.

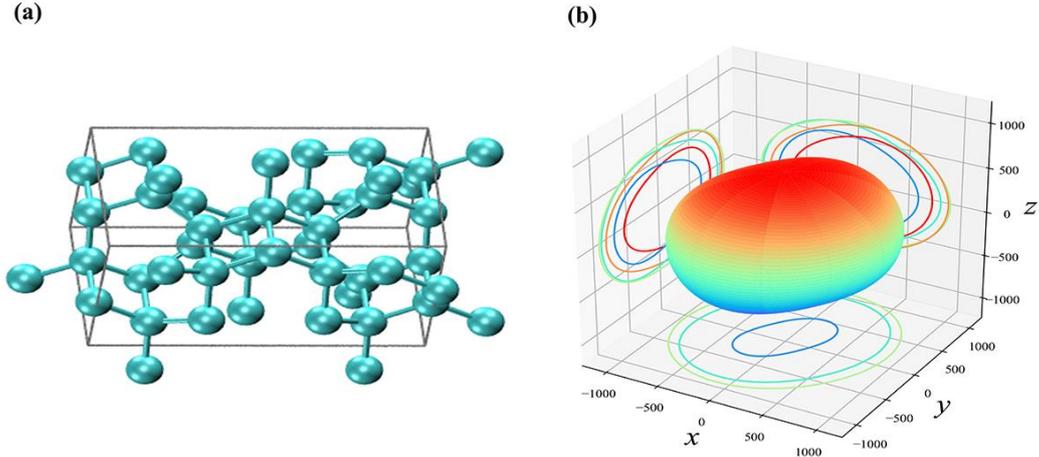

Figure 5. Crystal structure (a) and the directional dependence of the Young's modulus (b) of *Cmcm*-C24.

Table 1. Lattice constants (in Å), relative enthalpy to diamond (in eV/atom), shear modulus *G* (in GPa), Young's modulus *E* (in GPa), bulk modulus *B* (in GPa), and Poisson's ratio of *Cmcm*-C24.

| Structure | Space group | a | b | c | Relative enthalpy | G | E | B | Poisson's ratio |
|---|---|---|---|---|---|---|---|---|---|
| *Cmcm*-C24 | *Cmcm* | 7.641 | 4.361 | 4.459 | 0.32 | 382 | 859 | 379 | 0.12 |

To confirm the stability of the novel phase, the phonon spectra of the new carbon phase were calculated. The results (Figure 6(a)) show that *Cmcm*-C24 is thermodynamically stable because no imaginary frequencies are observed throughout the whole Brillouin zone.

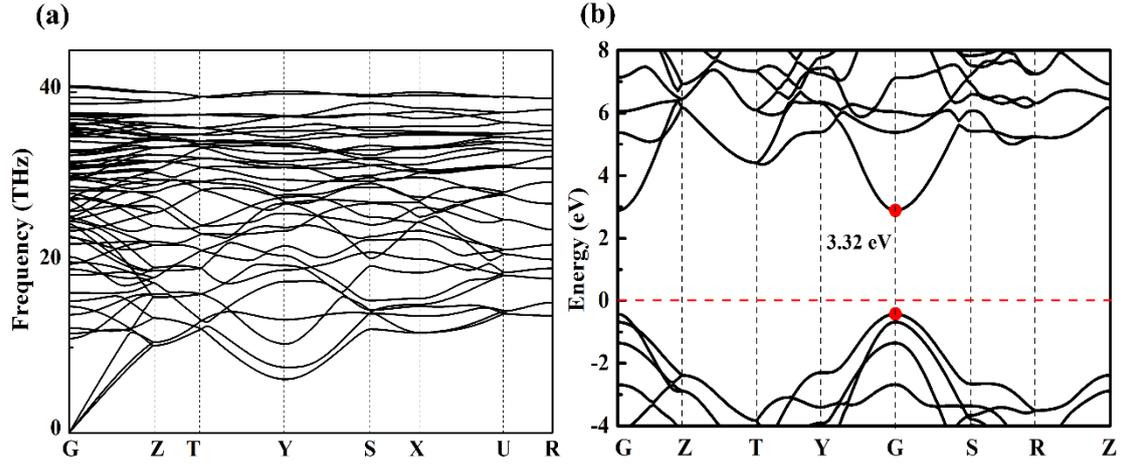

Figure 6. Phonon spectra (a) and electronic band structure at the HSE06 level (b) of the *Cmcm*-C24 phase.

The band structure of *Cmcm*-C24 was first calculated via standard DFT using the PBE functional, and the calculated bandgap was 2.23 eV. Because the standard DFT calculations usually underestimate bandgaps of semiconductors, the advanced, more computationally demanding hybrid functional HSE06 was used to predict the bandgap more accurately. The band structure of *Cmcm*-C24 at the HSE06 level is presented in Figure 6(b). The band structure clearly demonstrates that *Cmcm*-C24 exhibits a direct bandgap of 3.32 eV. In addition, the hardness of *Cmcm*-C24 was predicted to be 82 GPa by using the Lyakhov-Oganov approach [40], which is larger than the superhard standard of 40 GPa. Thus, this *Cmcm*-C24 is a potential superhard material.

## 4   Conclusion

In summary, we trained an ML model based on a self-built carbon structure database, and specifically predicted the elastic modulus of carbon materials. A comparison of the predictions with those of the AFLOW-ML model fully verified the validity and efficiency of our ML model on the basis of its application to predict the elastic modulus of carbon materials. In combination with the CALYPSO structure search software, our ML model was used instead of DFT calculations to predict the elastic modulus of the structure in the structure search process, thereby greatly reducing the time for calculating the elastic modulus. A new stable carbon allotrope, *Cmcm*-C24, with a low enthalpy and high elastic modulus was discovered through this improved structural search. Further, we confirmed that *Cmcm*-C24 is a direct-bandgap semiconductor with a gap of 3.32 eV at the HSE06 level.

**Acknowledgements**

This work was financially supported by the Fundamental Research Funds for the Central Universities, the Natural Science Foundation of China (Grant Nos.: 11965005 and 11964026), the 111 Project (B17035), and the Natural Science Basic Research plan in Shaanxi Province of China (Grant Nos.: 2020JM-186, 2020JM-621). All the authors thank the computing facilities at High Performance Computing Center of Xidian University.

**References**

[1] V. L. Deringer, G. Csányi, D. M. Proserpio, Extracting Crystal Chemistry from Amorphous Carbon Structures, *ChemPhysChem* 18, 8 (2017).

[2] Y. Zhuo, A. M. Tehrani, J. Brgoch, Predicting the Band Gaps of Inorganic Solids by Machine Learning, *J. Phys. Chem. Lett.* 9, 1668-1673 (2018).

[3] J. Lee, A. Seko, K. Shitara, K. Nakayama, I. Tanaka, Prediction model of band gap for inorganic compounds by combination of density functional theory calculations and machine learning techniques, *Phys. Rev. B* 93, 115104 (2016).

[4] P. Dey, J. Bible, S. Datta, S. Broderick, J. Jasinski, M. Sunkara, M. Menon, K. Rajan, Informatics-aided bandgap engineering for solar materials, *Comput. Mater. Sci.* 83, 185-195 (2014).

[5] A. O. Oliynyk, L. A. Adutwum, B. W. Rudyk, H. Pisavadia, S. Lotfi, V. Hlukhyy, J. J. Harynuk, A. Mar, J. Brgoch, Disentangling structural confusion through machine learning: Structure prediction and polymorphism of equiatomic ternary phases *ABC*, *J. Am. Chem. Soc.* 139, 17870-17881 (2017).

[6] A. O. Oliynyk, L. A. Adutwum, J. J. Harynuk, A. Mar, Classifying Crystal Structures of Binary Compounds AB through Cluster Resolution Feature Selection and Support Vector Machine Analysis. *Chem. Mater.* 28, 6672-6681 (2016).

[7] F. Legrain, J. Carrete, A. Roekeghem, S. Curtarolo, N. Mingo, How the Chemical Composition Alone Can Predict Vibrational Free Energies and Entropies of Solids. *Chem. Mater.* 29, 6220-6227 (2017).

[8] G. Pilania, P. V. Balachandran, C. Kim, T. Lookman, Finding New Perovskite Halides via Machine Learning, *Front. Mater*. 3, 19 (2016).

[9] O. Isayev, C. Oses, C. Toher, E. Gossett, S. Curtarolo, A. Tropsha, Universal fragment descriptors for predicting properties of inorganic crystals, *Nat. Commun.* 8, 15679 (2017).

[10] Y. Zhang, C. Ling, A strategy to apply machine learning to small datasets in materials science, *npj Comput. Mater.* 4, 25 (2018).

[11] A. M. Tehrani, A. O. Oliynyk, M. Parry, Z. Rizvi, S. Couper, F. Lin, L. Miyagi, T. D. Sparks, J. Brgoch, Machine Learning Directed Search for Ultraincompressible, Superhard Materials. *J. Am. Chem. Soc.* 140, 9844-9853 (2018).

[12] Y. W. Zhang, H. Wang, Y. C. Wang, L. J. Zhang, Y. M. Ma, Computer-assisted inverse design of inorganic electrides, *Phys. Rev. X* 7, 011017 (2017).

[13] X. X. Zhang, Y. C. Wang, J. Lv, C. Y. Zhu, Q. Li, M. Zhang, Q. Li, Y. M. Ma, First-principles structural design of superhard materials, *J. Chem. Phys.* 138, 114101 (2013).

[14] Y. C. Wang, J. Lv, L. Zhu, Y. M. Ma, CALYPSO: A method for crystal structure prediction, *Comput. Phys. Commun.* 183, 2063-2070 (2012).

[15] Y. Sun, J. Lv, Y. Xie, H. Y. Liu, Y. M. Ma, Route to a Superconducting Phase above Room Temperature in Electron-Doped Hydride Compounds under High Pressure, *Phys. Rev. Lett.* 123, 097001 (2019).

[16] Q. Wei, Q. Zhang, M. Zhang, H. Yan, L. Guo, B. Wei, A novel hybrid sp-$sp^2$ metallic carbon allorope. *Front. Phys.* 13, 136105 (2018).

[17] H. Yan, Z. Wei, M. Zhang, Q. Wei. Exploration of stable stoichiometries,

ground-state structures, and mechanical properties of the W-Si system, *Ceram. Int.* 46, 17034-17043 (2020).

[18] J. Lin, Z. Y. Zhao, C. Y. Liu, J. Zhang, X. Du, G. C. Yang, Y. M. Ma, $IrF_8$ Molecular Crystal under High Pressure, *J. Am. Chem. Soc.* 141, 5409-5414 (2019).

[19] Z. Y. Zhao, S. T. Zhang, T. Yu, H. Y. Xu, A. Bergara, G. C. Yang, Predicted Pressure-Induced Superconducting Transition in Electride $Li_6P$, *Phys. Rev. Lett.* 122, 097002 (2019).

[20] Q. Wei, W. Tong, R. K. Yang, H. Y. Yan, B. Wei, M. G. Zhang, X. C. Yang, R. Zhang, Orthorhombic C10: A new superdense carbon allotrope, *Phys. Lett. A* 383, 125861 (2019).

[21] Q. C. Tong, L. T. Xue, J. Lv, Y. C. Wang, Y. M. Ma, Accelerating CALYPSO structure prediction by data-driven learning of a potential energy surface, *Faraday Discuss.* 211, 31-43 (2018).

[22] K. Xia, H. Gao, C. Liu, J. N. Yuan, J. Sun, H. -T. Wang, D. Y. Xing, A novel superhard tungsten nitride predicted by machine-learning accelerated crystal structure search, *Sci. Bull.* 63, 817-824 (2018).

[23] R. Hoffmann, A. A. Kabanov, A. A. Golov, D. M. Proserpio, *Homo Citans* and Carbon Allotropes: For an Ethics of Citation, *Angew. Chem. -Int. Edit.* 55, 10962-10976 (2016).

[24] M. Gajdos, K. Hummer, G. Kresse, J. Furthmuller, F. Bechstedt, Linear optical properties in the projector-augmented wave methodology, *Phys. Rev. B* 73, 045112 (2006).

[25] G. Kresse, J. Furthmuller, Efficient iterative schemes for ab initio total-energy calculations using a plane-wave basis set, *Phys. Rev. B* 54, 11169 (1996).

[26] M. G. Zhang, H. Y. Yan, Q. Wei, Unexpected ground-state crystal structures and mechanical properties of transition metal pernitrides $MN_2$ (M= Ti, Zr, and Hf), *J. Alloy. Compd.* 774, 918-925 (2019).

[27] J. P. Perdew, K. Burke, M. Ernzerhof, Generalized Gradient Approximation Made Simple [Phys. Rev. Lett. 77, 3865 (1996)]. *Phys. Rev. Lett.* 78, 1396 (1997).

[28] J. P. Perdew, K. Burke, M. Ernzerhof, Generalized Gradient Approximation Made Simple. *Phys. Rev. Lett.* 77, 3865 (1997).

[29] A. Togo, F. Oba, I. Tanaka, First-principles calculations of the ferroelastic transition between rutile-type and $CaCl_2$-type $SiO_2$ at high pressures, *Phys. Rev. B* 78, 134106 (2008).

[30] L. Ward, A. Dunn, A. Faghaninia, N. E. R. Zimmermann, S. Bajaj, Q. Wang, et al., Matminer: An open source toolkit for materials data mining, *Comput. Mater. Sci.* 152, 60-69 (2018).

[31] E. Gossett, C. Toher, C. Oses, O. Isayev, F. Legrain, F. Rose, et al., AFLOW-ML: A RESTful API for machine-learning predictions of materials properties, *Comput. Mater. Sci.* 152, 134-145 (2018).

[32] A. R. Supka, T. E. Lyons, L. Liyanage, P. DAmico, R. A. R. A. Orabi, S. Mahatara, et al., AFLOW π: A minimalist approach to high-throughput ab initio calculations including the generation of tight-binding hamiltonians, *Comput. Mater. Sci.* 136, 76-84 (2017).

[33] M. J. Mehl, D. Hicks, C. Toher, O. Levy, R. M. Hanson, G. Hart, et al., The AFLOW Library of Crystallographic Prototypes: Part 1, *Comput. Mater. Sci.* 136, S1-S828 (2017).

[34] W. L. Mao, H. -K. Mao , P. J. Eng, T. P. Trainor, M. Newville, C. -C. Kao , et al., Bonding Changes in Compressed Superhard Graphite, *Science* 302, 425-427 (2003).

[35] Y. J Wang, J. E. Panzik, B. Kiefer, K. K. M. Lee, Crystal structure of graphite under room-temperature compression and decompression, *Sci. Rep.* 2, 520 (2012).

[36] Q. Li, Y. M. Ma, A. R. Oganov, H. B. Wang, H. Wang, Y. Xu, T. Cui, H. -K. Mao, G. G. Zou, Superhard Monoclinic Polymorph of Carbon, *Phys. Rev. Lett.* 102, 175506 (2009).

[37] E. Stavrou, S. Lobanov, H. F. Dong, A. R. Oganov, V. B. Prakapenka, Z. Konopkovaa, A. F. Goncharov, Synthesis of ultra-incompressible $sp^3$-hybridized carbon nitride with 1:1 stoichiometry, *Chem. Mater.* 2,8 6925-6933 (2016).

[38] M. Zhang, H. Liu, Q. Li, B. Gao, Y. C. Wang, H. D. Li, C. F. Chen, Y. M. Ma, Superhard $BC_3$ in cubic diamond structure, *Phys. Rev. Lett.* 114, 015502 (2015).

[39] R. Hill, The elastic behaviour of a crystalline aggregate, *Proc. Phys. Soc. London* 65, 349-354 (1952).

[40] A. Lyakhov, A. Oganov, Evolutionary search for superhard materials: Methodology and applications to forms of carbon and $TiO_2$, *Phys. Rev. B* 84, 092103 (2011).